\documentclass[journal,10pt]{IEEEtran}

\newtheorem{thm}{Theorem}[section]
\newtheorem{definition}[thm]{Definition}
\newtheorem{lemma}[thm]{Lemma}
\newtheorem{rem}[thm]{Remark}
\newtheorem{proposition}[thm]{Proposition}

\newtheorem{problem}[thm]{Problem}
\newtheorem{algorithm}[thm]{Algorithm}

\IEEEoverridecommandlockouts                              
\usepackage{amsfonts}
\usepackage{amsmath}
\usepackage{graphicx}      
\usepackage{float} 
\usepackage{psfrag} 
\usepackage{url}
\usepackage{subfig}
\usepackage{cite}

\title{\LARGE \bf Subspace identification of large-scale interconnected systems}

\author{Aleksandar Haber and Michel Verhaegen
\thanks{This research is supported by the Dutch Ministry of Economic Affairs and the Provinces of Noord-Brabant and Limburg in the frame of the "Pieken in de Delta" program.}
\thanks{A. Haber and M. Verhaegen are with Delft Center for Systems and Control, Delft University of Technology, Delft, 2628 CD, The Netherlands, (e-mail: {\tt\small a.haber@tudelft.nl}; {\tt\small m.verhaegen@tudelft.nl}).}
}
\begin{document}
\maketitle
\thispagestyle{empty}
\pagestyle{empty}
\maketitle
\begin{abstract}
We propose a decentralized subspace algorithm for identification of large-scale, interconnected systems that are described by sparse (multi) banded state-space matrices. First, we prove that the state of a local subsystem can be approximated by a linear combination of inputs and outputs of the local subsystems that are in its neighborhood. Furthermore, we prove that for interconnected systems with well-conditioned, finite-time observability Gramians (or observability matrices), the size of this neighborhood is relatively small. On the basis of these results, we develop a subspace identification algorithm that identifies a state-space model of a local subsystem from the local input-output data. Consequently, the developed algorithm is computationally feasible for interconnected systems with a large number of local subsystems. Numerical results confirm the effectiveness of the new identification algorithm. 
\end{abstract}
\section{Introduction}
Large-scale interconnected systems consist of a large number of local subsystems that are distributed and interconnected in a spatial domain \cite{benner2004,siljak1991,bamieh02distributedcontrol,dandrea2003}. The classical identification methods, such as the Subspace Identification Methods (SIMs) \cite{verhaegen2007} or Prediction Error Methods (PEMs) \cite{lennart1999system}, are not suitable for identification of large-scale interconnected systems. This is mainly because the computational and memory complexity of these methods scale at least with $O(N^{3})$ and $O(N^{2})$, respectively, where $N$ is the number of local subsystems of an interconnected system. Furthermore, the SIMs identify a state-space representation of an interconnected system, in which the interconnection structure is destroyed by unknown similarity transformation \cite{verhaegen2007}. However, for efficient distributed controller synthesis we need structured state-space models of interconnected systems \cite{paolo2009paper,bamieh02distributedcontrol,dandrea2003,justin2010distributed}.
\\
 On the other hand, the SIMs and the PEMs are centralized identification methods.
That is, these methods can be applied only if input-output data of all local subsystems can be collected and processed in one computing unit. In cases in which a large number of local sensors collect measurement data of local subsystems, it might not be possible to transfer local measurements to one centralized computing unit \cite{Rabbat2004}. In such situations, identification should be performed in a decentralized/distributed manner on a network of local computing units that communicate locally.  
\\
In \cite{paolo2009ident,massioniCirculant}, SIMs have been proposed for some special classes of interconnected systems. However, these identification algorithms are computationally infeasible for interconnected systems with a very large number of local subsystems.
\par
In this paper we propose a decentralized subspace algorithm for identification of large-scale, interconnected systems that are described by sparse (multi) banded state-space matrices. For example, these state-space models originate from discretization of 2D and 3D Partial Differential Equations (PDEs) \cite{benner2004,haberThesis}. First, we prove that the state of a local subsystem can be approximated by a linear combination of inputs and outputs of the local subsystems that are in its neighborhood.  Furthermore, we prove that for interconnected systems with well-conditioned, finite-time observability Gramians (or observability matrices), the size of this neighborhood is relatively small. On the basis of these results, we develop a subspace identification algorithm that identifies a state-space model of a local subsystem using only local input-output data. Consequently, the developed algorithm is computationally feasible for interconnected systems with an extremely large number of local subsystems (provided that the finite-time observability Gramian is well conditioned). We numerically illustrate the effectiveness of the new identification algorithm. 
\\
The paper is organized as follows. In Section \ref{problemFormulation}, we present the problem formulation. In Section \ref{mainTheorems}, we present theorems that are used in Section \ref{identSection} to develop the identification algorithm. In Section \ref{numericalSection} we present the results of numerical simulations and in Section \ref{sectionConclusion} we draw conclusions.
\section{Problem formulation}
\label{problemFormulation}
We briefly explain the notation used in this paper. A block diagonal matrix $X$ with blocks $X_{1},\ldots,X_{N}$ is denoted by $X=\text{diag}(X_{1},\ldots,X_{N})$. A column vector $\mathbf{z}=[\mathbf{z}_{1}^{T},\ldots, \mathbf{z}_{M}^{T}]^{T}$ is denoted by $\mathbf{z}=\text{col}(\mathbf{z}_{1},\ldots,\mathbf{z}_{M})$.
For the sake of presentation clarity, we consider the following state-space model: 
\begin{small}
\begin{align}
\mathcal{S} \left\{ \begin{array} {rl}
\underline{\mathbf{x}}(k+1)&=\underline{A}\underline{\mathbf{x}}(k)+\underline{B}\underline{\mathbf{u}}(k) \\
\underline{\mathbf{y}}(k)&=\underline{C}\underline{\mathbf{x}}(k)+\underline{\mathbf{n}}(k) \end{array} \right. \label{globalSys}
\end{align}
\end{small}
\begin{small}
\begin{align}
&\underline{A}= \left[\begin{array}{l}
 A_{1,1} \; \; E_{1,2}  \\
E_{2,1} \; \;  A_{2,2} \; \; E_{2,3}    \\
 \;\;\;\;\;\;\; \;\;\;\;\; \; \ddots   \\
 \;\;\;\;\;\;\;\; E_{i,i-1} \;\; A_{i,i}  \;\; E_{i,i+1}  \\
  \;\;\; \;\;\;\;\;\;\;\;\;\; \;\;\;\;\;\;\;\;\;\;\;\;\;\; \ddots   \\
   \;\;\;\; \;\;\;\; \;\;\;\;\;\ E_{N-1,N-2} \;\; A_{N-1,N-1}  \;\; E_{N-1,N}  \\
  \;\;\;\; \;\;\;\;\;\;\;\;\;\; \;\;\;\;\; \;\;\; \;\;\;\;\; \;\;\; \;\;\;\;\; \;\;\; E_{N,N-1} \;\;  A_{N,N} \end{array}\right]
\label{explanationGlobSys}
\end{align}
\end{small}
\begin{small}
\begin{align}
& \underline{B}= \text{diag} (B_{1},\ldots,B_{N}),\;\; 
\underline{C}= \text{diag} (C_{1},\ldots,C_{N}) \notag \\
&\underline{\mathbf{y}}(k)=\text{col}(\mathbf{y}_{1}(k),\ldots,\mathbf{y}_{N}(k))
,\;\; \underline{\mathbf{x}}(k)=\text{col}(\mathbf{x}_{1}(k),\ldots,\mathbf{x}_{N}(k))
\notag \\
&\underline{\mathbf{u}}(k)=\text{col}(\mathbf{u}_{1}(k),\ldots,\mathbf{u}_{N}(k))
,\;\;
\underline{\mathbf{n}}(k)=\text{col}(\mathbf{n}_{1}(k),\ldots,\mathbf{n}_{N}(k)) \notag
\end{align}
\end{small}
The system $\mathcal{S}$ is referred to as \textit{the global system}, with \textit{the global state} $\underline{\mathbf{x}}(k)\in \mathbb{R}^{Nn}$, \textit{the global input} $\underline{\mathbf{u}}(k) \in \mathbb{R}^{Nm} $, \textit{the global measured output} $\underline{\mathbf{y}}(k)\in \mathbb{R}^{Nr}$ and \textit{the global measurement noise} $\underline{\mathbf{n}}(k)\in \mathbb{R}^{Nr}$. The system matrices $\underline{A}\in \mathbb{R}^{Nn\times Nn}$, $\underline{B}\in  \mathbb{R}^{Nn\times Nm} $ and $\underline{C}\in  \mathbb{R}^{Nr\times Nn}$ are referred to as \textit{the global system matrices}. The global system $\mathcal{S}$ is an interconnection of $N$ \textit{local subsystems} $\mathcal{S}_{i}$: 
\begin{small}
\begin{align}
\mathcal{S}_{i} \left\{ \begin{array} {rll}
\mathbf{x}_{i}(k+1)&=A_{i,i}\mathbf{x}_{i}(k)+E_{i,i-1}\mathbf{x}_{i-1}(k)+E_{i,i+1}\mathbf{x}_{i+1}(k) \\ &+B_{i}\mathbf{u}_{i}(k)  \\ 
\mathbf{y}_{i}(k)&=C_{i}\mathbf{x}_{i}(k)+ \mathbf{n}_{i}(k)   \end{array} \right.
\label{localSubSys}
\end{align}
\end{small}
where $\mathbf{x}_{i}(k)\in \mathbb{R}^{n}$ is \textit{the local state} of the local subsystem $\mathcal{S}_{i}$, $\mathbf{x}_{i-1}(k)\in \mathbb{R}^{n}$ and $\mathbf{x}_{i+1}(k)\in \mathbb{R}^{n}$ are the local states of \textit{the neighboring local subsystems} $\mathcal{S}_{i-1}$ and $\mathcal{S}_{i+1}$ respectively, $\mathbf{u}_{i}(k)\in \mathbb{R}^{m}$ is \textit{the local input}, $\mathbf{y}_{i}(k)\in \mathbb{R}^{r}$ is \textit{the local measured output}, $\mathbf{n}_{i}(k)\in \mathbb{R}^{r}$ is \textit{the local measurement noise}. Without loss of generality, we assume that all local subsystems have identical local order $n\ll N$. All matrices in \eqref{localSubSys} are constant matrices and are referred to as \textit{the local system matrices}. In \eqref{localSubSys}, the index $i$ is referred to as \textit{the spatial index}. The spatial index takes the values from \textit{the spatial domain}  $\Pi=\{1,\ldots, N\}$. For example, the global state-space model \eqref{globalSys} can be obtained by discretizing the 2D heat equation using the finite difference method \cite{benner2004,haberThesis}. The identification algorithm proposed in this paper can be generalized to large-scale interconnected systems with sparse (multi) banded system matrices (see Remark \ref{remarkGeneralization}).
\par
 The set of local subsystems $V_{h}(S_{i})=\{\mathcal{S}_{i-h},\ldots,\mathcal{S}_{i+h}\}$ will be referred to as \textit{the neighborhood of $\mathcal{S}_{i}$}. \\
\begin{definition}
\label{structurePreservingSimDef}
\textit{The structure preserving similarity transformation} is the matrix $\underline{Q}=\text{diag}(Q_{1},\ldots, Q_{N})$, that transforms the global state-space model \eqref{globalSys}-\eqref{explanationGlobSys} into the following state-space model:
\begin{small}
\begin{align}
\hat{\mathcal{S}} \left\{ \begin{array} {rl}
\hat{\underline{\mathbf{x}}}(k+1)&=\underline{\hat{A}}\hat{\underline{\mathbf{x}}}(k)+\hat{\underline{B}}\underline{\mathbf{u}}(k) \\
\underline{\mathbf{y}}(k)&=\hat{\underline{C}}\hat{\underline{\mathbf{x}}}(k)+\underline{\mathbf{n}}(k) \end{array} \right. \label{globalSysHat}
\end{align}
\end{small}
where $\underline{\mathbf{x}}(k)=\underline{Q}\hat{\underline{\mathbf{x}}}(k)$, $\underline{\hat{A}}=\underline{Q}^{-1} \underline{A}\underline{Q}$, $\hat{\underline{B}}=\underline{Q}^{-1} \underline{B}$ and $\hat{\underline{C}}= \underline{C}\underline{Q}$. The matrix $\underline{\hat{A}}$ has block bandwidth equal to 1 (the same sparsity pattern like $\underline{A}$), while $\hat{\underline{B}}$ and $\hat{\underline{C}}$ are block diagonal matrices.\\
\end{definition} 
\begin{problem} \textit{Identification problem} 
\par
Consider the global system \eqref{globalSys} that consists of the interconnection of $N$ local subsystems \eqref{localSubSys}. Then, using the sequence of the global input-output data $\{\underline{\mathbf{y}}(k),\underline{\mathbf{u}}(k) \} $,
\begin{enumerate}
\item Estimate the order of the local subsystems $n$.
\item Identify the global state-space model \eqref{globalSys} up to a structure preserving similarity transformation.
\end{enumerate} 
\label{identificationProblem}
\end{problem}
\section{Main theorems}
\label{mainTheorems}
Starting from $k-p$ and by lifting \eqref{globalSys} $p$ time steps, we obtain:
\begin{small}
\begin{align}
\mathbf{Y}_{k-p}^{k}&=O_{p}\underline{\mathbf{x}}(k-p)+G_{p-1} \mathbf{U}_{k-p}^{k-1}+\mathbf{N}_{k-p}^{k} \label{liftedClassicalOutput}
\end{align}
\end{small}
where $\mathbf{Y}_{k-p}^{k}=\text{col}\left(\underline{\mathbf{y}}(k-p),\ldots,\underline{\mathbf{y}}(k)\right)$, $\mathbf{U}_{k-p}^{k-1}=\text{col}\left(\underline{\mathbf{u}}(k-p),\ldots,\underline{\mathbf{u}}(k-1)\right)$ and similarly we define $\mathbf{N}_{k-p}^{k}$. The matrix $O_{p}\in \mathbb{R}^{N(p+1)r\times Nn}$ is \textit{the $p$-steps observability matrix} and the matrix $G_{p-1}\in \mathbb{R}^{N(p+1)r\times Npm}$ is \textit{the $p-1$ steps impulse response matrix}\cite{verhaegen2007}. \\
On the other hand, from the state equation of the global state-space model \eqref{globalSys}, we have:
\begin{small}
\begin{align}
\mathbf{x}(k)&=\underline{A}^{p}\underline{\mathbf{x}}(k-p)+R_{p-1}
\mathbf{U}_{k-p}^{k-1} \label{liftedClassicalState}
\end{align}
\end{small}
where $R_{p-1}\in \mathbb{R}^{Nn\times Npm} $ is \textit{the $p-1$ steps controllability matrix}. \textit{The parameter $p$ should be selected such that it is much smaller than $N$ ($p\ll N$)}. This ensures that all lifted system matrices in \eqref{liftedClassicalOutput} and \eqref{liftedClassicalState} are sparse matrices.
\par
 Formation of the lifted equations \eqref{liftedClassicalOutput} and \eqref{liftedClassicalState} is the standard step in the classical SIMs. However, in this paper we use a different lifting technique. \textit{This lifting technique consists of first lifting the local outputs over the time domain and then lifting such lifted outputs over the spatial domain.} To explain this new lifting technique, we define the vector $\mathcal{Y}_{i,k-p}^{k}\in \mathbb{R}^{(p+1)r}$ as follows $\mathcal{Y}_{i,k-p}^{k}=\text{col}(\mathbf{y}_{i}(k-p),\mathbf{y}_{i}(k-p+1),\ldots,\mathbf{y}_{i}(k))$. In the same manner we define the lifted input vector $\mathcal{U}_{i,k-p}^{k-1}\in \mathbb{R}^{pm}$ and the lifted measurement noise vector $\mathcal{N}_{i,k-p}^{k}\in \mathbb{R}^{(p+1)r}$. Next, a column vector $\mathcal{Y}_{k-p}^{k}\in \mathbb{R}^{N(p+1)r}$ is defined as follows: $\mathcal{Y}_{k-p}^{k}=\text{col}(\mathcal{Y}_{1,k-p}^{k},\ldots,\mathcal{Y}_{N,k-p}^{k})$. In the same manner we define vectors $\mathcal{U}_{k-p}^{k-1}\in \mathbb{R}^{Npm}$ and $\mathcal{N}_{k-p}^{k}\in \mathbb{R}^{N(p+1)r}$. It can be easily proved that:
\begin{small}
\begin{align}
\mathcal{Y}_{k-p}^{k}=P_{Y}\mathbf{Y}_{k-p}^{k},\;\;\;\mathcal{N}_{k-p}^{k}=P_{Y}\mathbf{N}_{k-p}^{k},\;\;\; \mathcal{U}_{k-p}^{k-1}=P_{U}\mathbf{U}_{k-p}^{k-1}
\label{permutationMatrices}
\end{align}
\end{small}
where $P_{Y}$ and $P_{U}$ are permutation matrices. By multiplying the lifted equation \eqref{liftedClassicalOutput} from left with $P_{Y}$ and keeping in mind that permutation matrices are orthogonal, we obtain:
\begin{small}
\begin{align} 
\mathcal{Y}_{k-p}^{k}=\mathcal{O}_{p}\underline{\mathbf{x}}(k-p)+\mathcal{G}_{p-1}\mathcal{U}_{k-p}^{k-1}+\mathcal{N}_{k-p}^{k}
\label{liftedDataEq}
\end{align}
\end{small}
where the matrices $\mathcal{O}_{p} \in \mathbb{R}^{N(p+1)r\times Nn}$ and $\mathcal{G}_{p-1}\in \mathbb{R}^{N(p+1)r\times Npm}$ are defined as follows:
\begin{small}
\begin{align}
\mathcal{O}_{p}=P_{Y}O_{p},\;\; \mathcal{G}_{p-1}=P_{Y}G_{p-1}P_{U}^{T}
\label{explanationLiftedDataEq}
\end{align}
\end{small}
The equation \eqref{liftedDataEq} will be called \textit{the global data equation}. On the other hand, using the orthogonality of the permutation matrix $P_{U}$, from \eqref{liftedClassicalState} we have:
\begin{small}
\begin{align}
\underline{\mathbf{x}}(k)=\underline{A}^{p}\underline{\mathbf{x}}(k-p)+\mathcal{R}_{p-1}\mathcal{U}_{k-p}^{k-1} \label{liftedState}
\end{align}
\end{small}
where the matrix $\mathcal{R}_{p-1} \in \mathbb{R}^{Nn\times Npm}$ is defined by $\mathcal{R}_{p-1}=R_{p-1}P_{U}^{T}$. \textit{The matrix $\mathcal{O}_{p}$ is a sparse banded matrix, with the block bandwidth equal to $p$. Similarly, the matrices $\mathcal{G}_{p-1}$ and $\mathcal{R}_{p-1}$ are sparse banded matrices with the block bandwidth equal to $p-1$}. For more information about the structure of $\mathcal{O}_{p}$, $\mathcal{G}_{p-1}$, and $\mathcal{R}_{p-1}$, the interested reader is advised to consult Chapter 3 of \cite{haberThesis}.
\\
The following lemma will be used to prove Theorems \ref{stateEstimatorLemma} and \ref{mainTheoremSpatialDecay}. \\
\begin{lemma}
Assume that $p\ge \nu$, where $\nu$ is the observability index of the global system \cite{luenberger1971,haber2013mhe}. Then $\text{rank}(\mathcal{O}_{p})=nN$.
\label{observabilityTheorem}  \\
\end{lemma}
\textit{Proof}. 
The matrix $\mathcal{O}_{p}$ is defined by permuting the rows of $O_{p}$. Since this permutation does not change the rank of $O_{p}$, we have: $\text{rank}(O_{p})=\text{rank}(\mathcal{O}_{p})$. Now, let $p \ge \nu$, where $\nu$ is the observability index of the global system. This implies $\text{rank}(O_{p})=\text{rank}(\mathcal{O}_{p})=nN$.  $\hfill \square$
\\ \par
Throughout the reminder of the paper, the matrix $\mathcal{O}_{p}$ will be called \textit{the structured observability matrix} of the global system \eqref{globalSys}. \textit{The finite-time observability Gramian} of the global system \eqref{globalSys}, denoted by $\mathcal{J}_{2p}\in \mathbb{R}^{Nn\times Nn}$, can be expressed as follows \cite{lim1996hankel,antoulas2005approximation}:
\begin{small}
\begin{align}
\mathcal{J}_{2p}=O_{p}^{T}O_{p}
\label{obsGramianGlobal}
\end{align}
\end{small} 
From \eqref{explanationLiftedDataEq} and \eqref{obsGramianGlobal}, we have:
\begin{small}
\begin{align}
\mathcal{J}_{2p}=\mathcal{O}_{p}^{T}P_{Y}P_{Y}^{T}\mathcal{O}_{p}=\mathcal{O}_{p}^{T}\mathcal{O}_{p}
\label{FFmatrix}
\end{align}
\end{small}
Taking into account that $\mathcal{O}_{p}$ has block bandwidth equal to $p$, from \eqref{FFmatrix} we have that $\mathcal{J}_{2p}$ is a sparse banded matrix with the block bandwidth equal to $2p$. Throughout the remainder of the paper, the condition number of $\mathcal{J}_{2p}$ will be denoted by $\kappa$. The inverse of $\mathcal{J}_{2p}$ will be denoted by $\mathcal{D}\in \mathbb{R}^{Nn\times Nn}$. For the sequel we will assume that the matrix $\mathcal{J}_{2p}$ is well-conditioned. \textit{As we will prove later, this assumption ensures that the local state of $\mathcal{S}_{i}$ can be approximated by a linear combination of input and output data of local subsystems that are in a relatively small neighborhood of $\mathcal{S}_{i}$}. \\
\begin{thm}
Let $p\ge \nu$, where $\nu$ is the observability index of the global system. Then,
\begin{small} 
\begin{align}
  \underline{\mathbf{x}}(k)=&\underline{A}^{p}\mathcal{D}\left(\mathcal{O}_{p}^{T}\mathcal{Y}_{k-p}^{k}-\mathcal{O}_{p}^{T}\mathcal{G}_{p-1}\mathcal{U}_{k-p}^{k-1}-\mathcal{O}_{p}^{T}\mathcal{N}_{k-p}^{k}\right)\notag \\&+\mathcal{R}_{p-1}\mathcal{U}_{k-p}^{k-1}
\label{expression4Final}
\end{align}
\end{small}
\label{stateEstimatorLemma}
where $\mathcal{D}=\mathcal{J}_{2p}^{-1}$.
\end{thm}
\textit{Proof}. From \eqref{liftedDataEq}, we have:
\begin{small}
\begin{align} 
\mathcal{O}_{p}\underline{\mathbf{x}}(k-p)=\mathcal{Y}_{k-p}^{k}-\mathcal{G}_{p-1}\mathcal{U}_{k-p}^{k-1}-\mathcal{N}_{k-p}^{k}
\label{expresion1}
\end{align}
\end{small}
Because $p\ge \nu$, from Lemma \ref{observabilityTheorem} we have: $\text{rank}(\mathcal{O}_{p})=Nn$. This implies that $\mathcal{J}_{2p}$ is positive definite and invertible. Because of this, from \eqref{expresion1} we have:
\begin{small}
\begin{align} 
\underline{\mathbf{x}}(k-p)=\mathcal{D}\left(\mathcal{O}_{p}^{T}\mathcal{Y}_{k-p}^{k}-\mathcal{O}_{p}^{T}\mathcal{G}_{p-1}\mathcal{U}_{k-p}^{k-1}-\mathcal{O}_{p}^{T}\mathcal{N}_{k-p}^{k}\right)
\label{expresion2}
\end{align}
\end{small}
where $\mathcal{D}=\mathcal{J}_{2p}^{-1}$. Substituting \eqref{expresion2} in  \eqref{liftedState} we arrive at \eqref{expression4Final}. $\hfill \square$
\par
 Although $\mathcal{J}_{2p}$ is a sparse banded matrix, in the general case, $\mathcal{D}$ is a dense matrix. However, by examining the entries of $\mathcal{D}$, it can be observed that the absolute values of the off-diagonal entries decay\footnote{The absolute values of off-diagonal elements can also oscillate. However, there should exist a decaying exponential function that bounds these oscillations.}, as they are further away from the main diagonal. This phenomena has been studied in \cite{benzi2007} and \cite{demko1984}, and it is illustrated in Fig. \ref{decayFinv}(a) in Section \ref{numericalSection}. Next, we define exponentially off-diagonally decaying matrices. \\
\begin{definition} \cite{benzi2007} We say that an $nN \times nN$ matrix $Z=[z_{i,j}]$ is an exponentially off-diagonally decaying matrix if there exist constants $c,\lambda\in\mathbb{R}$, $c>0$ and $\lambda\in(0,1)$, such that:
\begin{small}
\begin{align}
|z_{i,j}|\le c \lambda^{|i-j|}
\label{spatialDecayDefinition}
\end{align}
\end{small}
for all $i,j=1,\ldots,nN$. \\ \end{definition} 
\begin{thm} Let $p \ge \nu$, where $\nu$ is the observability index of the global system, and consider the finite-time observability Gramian $\mathcal{J}_{2p}$ and its inverse $\mathcal{D}$. The matrix $\mathcal{D}$ is an exponentially off-diagonally decaying matrix with:
\begin{small}
\begin{align}
 \lambda=\left(\frac{\sqrt{\kappa}-1}{\sqrt{\kappa}+1}\right)^{1/g} , \; c=\left\|\mathcal{D}\right\|_{2}\max \left\lbrace 1,\frac{(1+\sqrt{\kappa})^{2}}{2\kappa} \right\rbrace 
\label{constantsDefinition}
\end{align}
\end{small}
\label{mainTheoremSpatialDecay}
\end{thm}
where $g$ is the bandwidth\footnote{The bandwidth $g$ is defined by $g=m/2$, where $m$ is a constant in Eq. (2.6) in \cite{demko1984}.} of $\mathcal{J}_{2p}$ ($g$ is proportional to $p$).
\textit{Proof}. Because $p\ge \nu$ from Lemma \ref{observabilityTheorem} it follows that $\mathcal{O}_{p}$ has full column rank. This implies that $\mathcal{J}_{2p}$ is a symmetric positive definite matrix. Because of this, from \cite{demko1984} (see Theorem 2.4 and Proposition 2.2 in \cite{demko1984}, and for a more general proof, see \cite{benzi2007}), if follows that $\mathcal{D}$ is an exponentially off-diagonally decaying matrix with the
constants $c$ and $\lambda$ given by \eqref{constantsDefinition}. $\hfill \square$ \\

Because by assumption the matrix $\mathcal{J}_{2p}$ is well conditioned and $p\ll N$, from \eqref{constantsDefinition} it follows that the off-diagonal decay of $\mathcal{D}$ is rapid. Because of this, the matrix $\mathcal{D}$ can be approximated by a sparse banded matrix \cite{benzi2007}. \\
\begin{definition} \cite{benzi2007} Let $\mathcal{D}=[d_{i,j}]$. The matrix $\breve{\mathcal{D}}=[\breve{d}_{i,j}]$ with its elements defined by:
\begin{small}
\begin{align}
\breve{d}_{i,j}=\left\{ \begin{array}{ll}
    d_{i,j} \;\;\; \;\; \text{if} \; \; |i-j|\le s \textrm{,} \\
 \;\;\; 0  \;\; \;\;\;\; \text{if} \; \; |i-j|> s \textrm{,} 
   \end{array} \right.
\label{blockApproximation2}
\end{align}
\end{small}
is an approximation of $\mathcal{D}$.
\label{blockBandedDefinition} \\
\end{definition}
\begin{proposition}
Consider the matrix $\mathcal{D}\in \mathbb{R}^{nN\times nN}$ and its approximation $\breve{\mathcal{D}}\in \mathbb{R}^{nN\times nN} $. Then,
\begin{small}
\begin{align}
\left\|\mathcal{D}-\breve{\mathcal{D}}\right\|_{1} < c k_{1},\;\;\; k_{1}=2\lambda^{s+1} \frac{1-\lambda^{Nn-s}}{1-\lambda} 
\label{boundOnApproximation} 
\end{align}
\end{small}
where the constants $c$ and $\lambda$ are defined in \eqref{constantsDefinition}. Moreover, the parameter $k_{1}$ is an increasing function of $\kappa$.
\label{blockBandedApprox}
\end{proposition} 
\textit{Proof}. See the proof of Proposition 5.4 in \cite{haberThesis}.
$\hfill \square$ \\ \\
 Let us assume that the bandwidth of  $\breve{\mathcal{D}}$ is chosen such that $s=nt$, where $t$ is a positive integer. Similarly to the partitioning of the matrix $\underline{A}$, defined in \eqref{explanationGlobSys}, we partition $\breve{\mathcal{D}}$ into $N^{2}$ blocks, where each block is a matrix of dimension $n\times n$. After this partitioning, the matrix $\breve{\mathcal{D}}$ has the block bandwidth equal to $t$. Accordingly, throughout the remainder of the paper, the matrix $\breve{\mathcal{D}}$ will be denoted by $\breve{\mathcal{D}}_{t}$.
\\
From Proposition \ref{blockBandedApprox} we see that the approximation accuracy increases as $s$ increases or equivalently, as  $t$ increases. Furthermore, we see that the approximation accuracy is better when $\kappa$ is smaller. Because $\mathcal{J}_{2p}$ is well conditioned, there exists $s\ll nN$, or equivalently, $t\ll N$, for which the accuracy of approximating $\mathcal{D}$ by $\breve{\mathcal{D}}_{t}$ is relatively good \cite{benzi2007}. For the sequel we assume that $t\ll N$.
\par
By substituting $\mathcal{D}$ with $\breve{\mathcal{D}}_{t}$ in \eqref{expression4Final}, we define an approximate global state:
\begin{small} 
\begin{align}
 \breve{\underline{\mathbf{x}}}(k)=& \underline{A}^{p}\breve{\mathcal{D}}_{t}\left(\mathcal{O}_{p}^{T}\mathcal{Y}_{k-p}^{k}-\mathcal{O}_{p}^{T}\mathcal{G}_{p-1}\mathcal{U}_{k-p}^{k-1}-\mathcal{O}_{p}^{T}\mathcal{N}_{k-p}^{k}\right) +\mathcal{R}_{p-1}\mathcal{U}_{k-p}^{k-1}
\label{expresion32}
\end{align}
\end{small}
 For the sequel we will partition $\breve{\underline{\mathbf{x}}}(k)$ as follows: $\breve{\underline{\mathbf{x}}}(k)=\text{col}(\breve{\mathbf{x}}_{1}(k),\ldots,\breve{\mathbf{x}}_{N}(k))$, where $\breve{\mathbf{x}}_{i}(k)\in \mathbb{R}^{n}$,  $\forall i\in \Pi$. From \eqref{expresion32} we have that $\breve{\mathbf{x}}_{i}(k)$ is a linear combination of the lifted local inputs, lifted local outputs and lifted local measurement noises of the local subsystems belonging to the neighborhoods $V_{3p+t-1}\left(\mathcal{S}_{i}\right)$, $V_{2p+t}\left(\mathcal{S}_{i}\right)$ and $V_{2p+t}\left(\mathcal{S}_{i}\right)$, respectively. 
Because $t\ll N$, these neighborhoods are small. By substituting in \eqref{localSubSys} the local states $\mathbf{x}_{i-1}(k)$ and $\mathbf{x}_{i+1}(k)$ with their approximations, $\breve{\mathbf{x}}_{i-1}(k)$ and $\breve{\mathbf{x}}_{i+1}(k)$, we obtain the following approximate state-space model:
\begin{small}
\begin{align}
&\begin{array}{l}
\mathbf{x}_{i}(k+1) \approx A_{i,i}\mathbf{x}_{i}(k)+\breve{Q}_{i}\breve{\Omega}_{i}+\breve{B}_{i}^{(3)}\breve{N}_{i}^{(1)} \\
\mathbf{y}_{i}(k)=C_{i}\mathbf{x}_{i}(k)+ \mathbf{n}_{i}(k)  
\end{array}
\label{unifiedSys} \\
&\breve{Q}_{i}=\begin{bmatrix} \breve{B}_{i}^{(1)} & \breve{B}_{i}^{(2)} \end{bmatrix},\;
\breve{\Omega}_{i}=\begin{bmatrix} \breve{Y}_{i}^{(1)} \\ \breve{U}_{i}^{(2)}   \end{bmatrix},\; \notag \\
& \breve{Y}_{i}^{(1)}=\text{col}\left(\mathcal{Y}_{i-1-2p-t,k-p}^{k}, \ldots, \mathcal{Y}_{i+1+2p+t,k-p}^{k} \right),\; \notag \\
& \breve{U}_{i}^{(2)}=\text{col}\left( \mathcal{U}_{i-3p-t,k-p}^{k-1}, \ldots , \mathcal{U}_{i+3p+t,k-p}^{k-1} \right), \notag \\ &\breve{N}_{i}^{(1)}=\text{col}\left( \mathcal{N}_{i-1-2p-t,k-p}^{k}, \ldots ,\mathcal{N}_{i+1+2p+t,k-p}^{k} \right)
\label{finalIdentificationModel}
\end{align}
\end{small}
    \section{Identification algorithm}
\label{identSection}

The main idea of the identification algorithm it to use the approximate state-space model \eqref{unifiedSys} to estimate the state sequence of the local subsystem $\mathcal{S}_{i}$. This step has to be repeated for all $N$ local subsystems. Because $t\ll N$ and $p\ll N$, the input $\breve{\Omega}_{i}$ of \eqref{unifiedSys}, contains input-output data of local subsystems that are in a relatively small neighborhood of $\mathcal{S}_{i}$. Consequently, using the SIMs \cite{verhaegen2007} we can estimate the state sequence of \eqref{unifiedSys} in a computationally efficient manner. \textit{The computational complexity of estimating the state of \eqref{unifiedSys} is independent from the total number of local subsystems $N$}. However, the problem lies in the fact that we do not know in advance the precise value of $t$ that determines the form of the input $\breve{\Omega}_{i}$. As it will be explained in Section \ref{commentsSubSection}, this problem can be solved by choosing several values of $t$ and by computing the Variance Accounted For (VAF) of the identified models.
\\
Let the estimated state sequence of the approximate state-space model \eqref{unifiedSys} be denoted by $\{\hat{\mathbf{x}}_{i}(k)\}$. The state sequence $\{\hat{\mathbf{x}}_{i}(k)\}$ is approximately related to the ``true" state sequence of the local subsystem $\mathcal{S}_{i}$, via the following transformation:
\begin{small}
\begin{align}
 \mathbf{x}_{i}(k) \approx Q_{i}\hat{\mathbf{x}}_{i}(k)
\label{similarityTransformation12}
\end{align}
\end{small}
where $Q_{i}$ is a square, invertible matrix. We will denote the estimated state-sequences of the local subsystems $\mathcal{S}_{i-1}$ and $\mathcal{S}_{i+1}$ (that are estimated on the basis of \eqref{unifiedSys}) by $\{\hat{\mathbf{x}}_{i-1}(k)\}$ and $\{\hat{\mathbf{x}}_{i+1}(k)\}$, respectively. Similarly to \eqref{similarityTransformation12}, we have:
\begin{small}
\begin{align}
 \mathbf{x}_{i-1}(k)\approx Q_{i-1}\hat{\mathbf{x}}_{i-1}(k),\; \mathbf{x}_{i+1}(k) \approx Q_{i+1}\hat{\mathbf{x}}_{i+1}(k)
\label{similarityTransformation11}
\end{align}
\end{small}
where $Q_{i-1}$ and $Q_{i+1}$ are invertible matrices. By substituting \eqref{similarityTransformation12} and \eqref{similarityTransformation11} in \eqref{localSubSys}, and by multiplying the state-equation with $Q_{i}^{-1}$, we obtain:
\begin{small}
\begin{align}
\begin{array}{l}
\hat{\mathcal{S}}_{i} \left\{ \begin{array} {rl}
\hat{\mathbf{x}}_{i}(k+1)& \approx \underbrace{Q_{i}^{-1}A_{i,i}Q_{i}}_{\hat{A}_{i,i}}\hat{\mathbf{x}}_{i}(k)+\underbrace{Q_{i}^{-1}E_{i,i-1}Q_{i-1}}_{\hat{E}_{i,i-1}}\hat{\mathbf{x}}_{i-1}(k) \\ & +\underbrace{Q_{i}^{-1}E_{i,i+1}Q_{i+1}}_{\hat{E}_{i,i+1}}\hat{\mathbf{x}}_{i+1}(k)  +\underbrace{Q_{i}^{-1}B_{i}}_{\hat{B}_{i}}\mathbf{u}_{i}(k) \\
\mathbf{y}_{i}(k)& \approx \underbrace{C_{i}Q_{i}}_{\hat{C}_{i}}\hat{\mathbf{x}}_{i}(k)+ \mathbf{n}_{i}(k)   \end{array} \right. 
\end{array}
\label{localSubSys4}
\end{align}
\end{small}
State-space model \eqref{localSubSys4} tells us that once the local state sequences are estimated, the local system matrices $\{\hat{A}_{i,i}, \hat{E}_{i,i-1}, \hat{E}_{i,i+1}, \hat{B}_{i}, \hat{C}_{i}\}$ can be estimated by solving a least-squares problem formed on the basis of:
\begin{small}
\begin{align}
&\left[\begin{array}{l}\hat{\mathbf{x}}_{i}(k+1) \;\;\; \mathbf{y}_{i}(k) \end{array}\right]\approx \notag \\
& \underbrace{\left[\begin{array}{l} \hat{A}_{i,i} \;\; \hat{E}_{i,i-1}\;\;\hat{E}_{i,i+1}\;\; \hat{B}_{i}\;\; \hat{C}_{i}\end{array}\right]}_{\text{matrices to be estimated}}\begin{bmatrix}\hat{\mathbf{x}}_{i}(k) & 0 \\ \hat{\mathbf{x}}_{i-1}(k) & 0 \\
\hat{\mathbf{x}}_{i+1}(k) & 0 \\ \mathbf{u}_{i}(k) & 0 \\ 0 & \hat{\mathbf{x}}_{i}(k) \end{bmatrix}+\begin{bmatrix}0 & \mathbf{n}_{i}(k) \end{bmatrix}
\label{simpleLinearRegression}
\end{align}
\end{small}
Using the same principle, we can estimate the local system matrices of other local subsystems. Using the estimates of the local system matrices, we can form the estimates $\{ \underline{\hat{A}},\underline{\hat{B}}, \underline{\hat{C}}\}$. Next, from \eqref{localSubSys4} we have:
\begin{small}
\begin{align}
&\hat{A}_{i,i}\approx Q_{i}^{-1}A_{i,i}Q_{i},\;\; \hat{E}_{i,i-1}\approx Q_{i}^{-1}E_{i,i-1}Q_{i-1}, \notag \\ & \hat{E}_{i,i+1}\approx Q_{i}^{-1}E_{i,i+1}Q_{i+1},\;\; 
\hat{B}_{i}\approx Q_{i}^{-1}B_{i},\;\; \hat{C}_{i}\approx C_{i}Q_{i}
\label{finalStrategy}
\end{align}
\end{small}
Since \eqref{localSubSys4} and \eqref{finalStrategy} hold for all $i\in \Pi$, we conclude that $\underline{\mathbf{x}}(k)\approx \underline{Q}\hat{\underline{\mathbf{x}}}(k)$, $\underline{\hat{A}}\approx \underline{Q}^{-1} \underline{A}\underline{Q}$, $\hat{\underline{B}}\approx \underline{Q}^{-1} \underline{B}$, and $\hat{\underline{C}}\approx \underline{C}\underline{Q}$,
where $\underline{Q}=\text{diag}(Q_{1},\ldots, Q_{N})$ is the structure preserving similarity transformation (see Definition \ref{structurePreservingSimDef}). This shows that the identified model is (approximately) similar to the global state-space model \eqref{globalSys}. We are now ready to formally state the identification algorithm.
\\
\begin{algorithm} \textit{Identification of the global state-space model}\\
For $i=1,\ldots,N$, perform the steps 1 and 2: \\ 
1. Choose the parameters $p$ and $t$ and form the input vector $\breve{\Omega}_{i}$  of the state space model \eqref{unifiedSys}.\\
 2. Estimate the local state sequence $\{\mathbf{x}_{i}(k)\}$ of the state space model \eqref{unifiedSys} using the SIM.\\
After the steps 1 and 2 are completed, the state sequences $\{\hat{\mathbf{x}}_{i}(k)\}$, $i=1,\ldots,N$, are available. For $i=1,\ldots,N$, perform the following step:\\
  3. On the basis of \eqref{simpleLinearRegression} form a least-squares problem, and estimate the local system matrices $\{\hat{A}_{i,i}, \hat{E}_{i,i-1}, \hat{E}_{i,i+1}, \hat{B}_{i}, \hat{C}_{i}\}$. \\
  4. Using the estimates $\{\hat{A}_{i,i}, \hat{E}_{i,i-1}, \hat{E}_{i,i+1}, \hat{B}_{i}, \hat{C}_{i}\}$, $i=1,\ldots,N$, form the global system matrices $\{ \underline{\hat{A}},\underline{\hat{B}}, \underline{\hat{C}}\}$.
\label{identificationAlgorithm}
\end{algorithm}
\subsection{Comments on the identification algorithm}
\label{commentsSubSection}
The theory developed in this paper predicts that for systems with well-conditioned, finite-time observability Gramians, there should exist $t\ll N$ for which the matrix $\mathcal{D}$ can be accurately approximated by the sparse banded matrix $\breve{\mathcal{D}}_{t}$. This implies that in the first step of Algorithm \ref{identificationAlgorithm}, we can select any $t$ that satisfies $t\ll N$.  
After that, the model should be identified and the VAF should be calculated. If the VAF value of the identified model is not high enough, then a new value of $t$ needs to be chosen and identification procedure needs to be repeated (usually the new value should be larger than the previous one). This has to be repeated until a relatively high value of the VAF of the identified model is reached. Because the local subsystems are not identical, it might happen that the rate of the off-diagonal decay of $\mathcal{D}$ varies from one row to another. This implies that the matrix $\mathcal{D}$ can be approximated by a banded matrix, which bandwidth is row dependent. That is, for each local subsystem $\mathcal{S}_{i}$, it is possible to find a different parameter $t_{i}$ that determines the input $\breve{\Omega}_{i}$.  
\\
As it is shown in the next section, the form of the input $\breve{\Omega}_{i}$ can cause ill-conditioning of the data  matrices used in the SIM. This is because $\breve{\Omega}_{i}$ consists of the delayed inputs and outputs of the local subsystems. Some of the outputs might be depending on the past local inputs and local outputs. This problem can be resolved either by regularizing the data matrices used in the SIM or by eliminating certain outputs and inputs from $\breve{\Omega}_{i}$. In this paper, we do not analyze the consistency of the identification algorithm. The consistency analysis is left for future research. The estimates of local subsystem matrices obtained using the proposed identification algorithm, can be used as initial guesses of the decision variables of a computationally efficient, parameter optimization method presented in Chapter 6 of \cite{haberThesis}. This way, the identification results can be additionally improved.
\begin{rem}
Algorithm \ref{identificationAlgorithm} can be generalized for global systems described by sparse, multi-banded, state-space matrices. For example, these systems originate from discretization of 3D PDEs using the finite difference method \cite{haberThesis}. Using the lifting technique presented in Section  \ref{mainTheorems}, it can be easily shown that the finite-time observability Gramian  $\mathcal{J}_{2p}$ of this class of systems, is a sparse, multi-banded matrix, for more details see chapters 2 and 3 of \cite{haberThesis}. In \cite{grote1997,benzi2007,haberThesis}, it has been shown that inverses  of sparse multi-banded matrices can be approximated by sparse multi-banded matrices. That is, the inverse of $\mathcal{J}_{2p}$ can be approximated by a sparse multi-banded matrix. From the identification point of view, this implies that the state of a local subsystem can be identified using the local input-output data of local subsystems that are in its neighborhood. Depending on the interconnection pattern of local subsystems, this neighborhood can be a 2D or a 3D neighborhood. 
\label{remarkGeneralization}
\end{rem}

\section{Numerical experiments}
\label{numericalSection}
The data generating model, is a global state-space model consisting of $N=500$ identical local subsystems. The local system matrices of each local subsystem are given by:
\begin{small}
\begin{align}
& A=\begin{bmatrix} 0.5728  & 0.1068   \\
                   0.1068  &  0.5728\end{bmatrix}, E=\begin{bmatrix}       0.1068      & 0 \\
         0  &     0.1068  \end{bmatrix},
 B=\begin{bmatrix} 0.2136\\  0.1068 \end{bmatrix}\notag \\ & C=\begin{bmatrix} 1 & 0 \end{bmatrix}
\label{modelIdentification}
\end{align}
\end{small}
The model \eqref{modelIdentification} is obtained using the finite-difference approximation of the heat equation, see Chapter 2 of \cite{haberThesis} (the thermal diffusivity constant is $0.6$, and the temporal and spatial discretization steps are $h=5$ and $L=5.3$, respectively). Every local input is a zero-mean Gaussian white noise. Every local output is corrupted by a zero-mean Gaussian white noise. Signal to Noise Ratio (SNR) of each local output is $25$ $[\text{dB}]$. In total 100 identification experiments are performed.
\par
For identification of the local state of \eqref{unifiedSys}, we use the SIM method summarized in \cite{haber2} (other SIMs can be also used). This SIM is a modified version of the SIM presented in \cite{jansson2003}. The SIM is applied with the past and future windows equal to $15$ and $10$, respectively.  Because all local subsystems have identical local system matrices \eqref{modelIdentification}, to identify the global state-space model we only need to perform three identification experiments. Namely, using \eqref{unifiedSys} we first estimate the state sequence of the local subsystem $\mathcal{S}_{2}$. Using the same methodology, we estimate the state sequence of $\mathcal{S}_{1}$. In the final identification step, we use the state-space model \eqref{simpleLinearRegression} (we set $i=1$ in \eqref{simpleLinearRegression}) and the sequences $\{\hat{\mathbf{x}}_{1}(k)\}$,$\{\hat{\mathbf{x}}_{2}(k)\}$ and $\{\mathbf{y}_{1}(k),\mathbf{u}_{1}(k)\}$, to form a least-squares problem. By solving this least-squares problem we estimate the local system matrices $\{\hat{A},\hat{E},\hat{B},\hat{C} \}$. Using $\{\hat{A},\hat{E},\hat{B},\hat{C} \}$ we form  $\{\underline{\hat{A}},\underline{\hat{B}},\underline{\hat{C}}\}$ and we compute the VAF of the identified model. 
\par
\begin{figure}[H]
\centering
 \subfloat[]{\includegraphics[width=4.4cm, height=3.9cm, trim=0mm 0mm 0mm 0mm,clip=true]{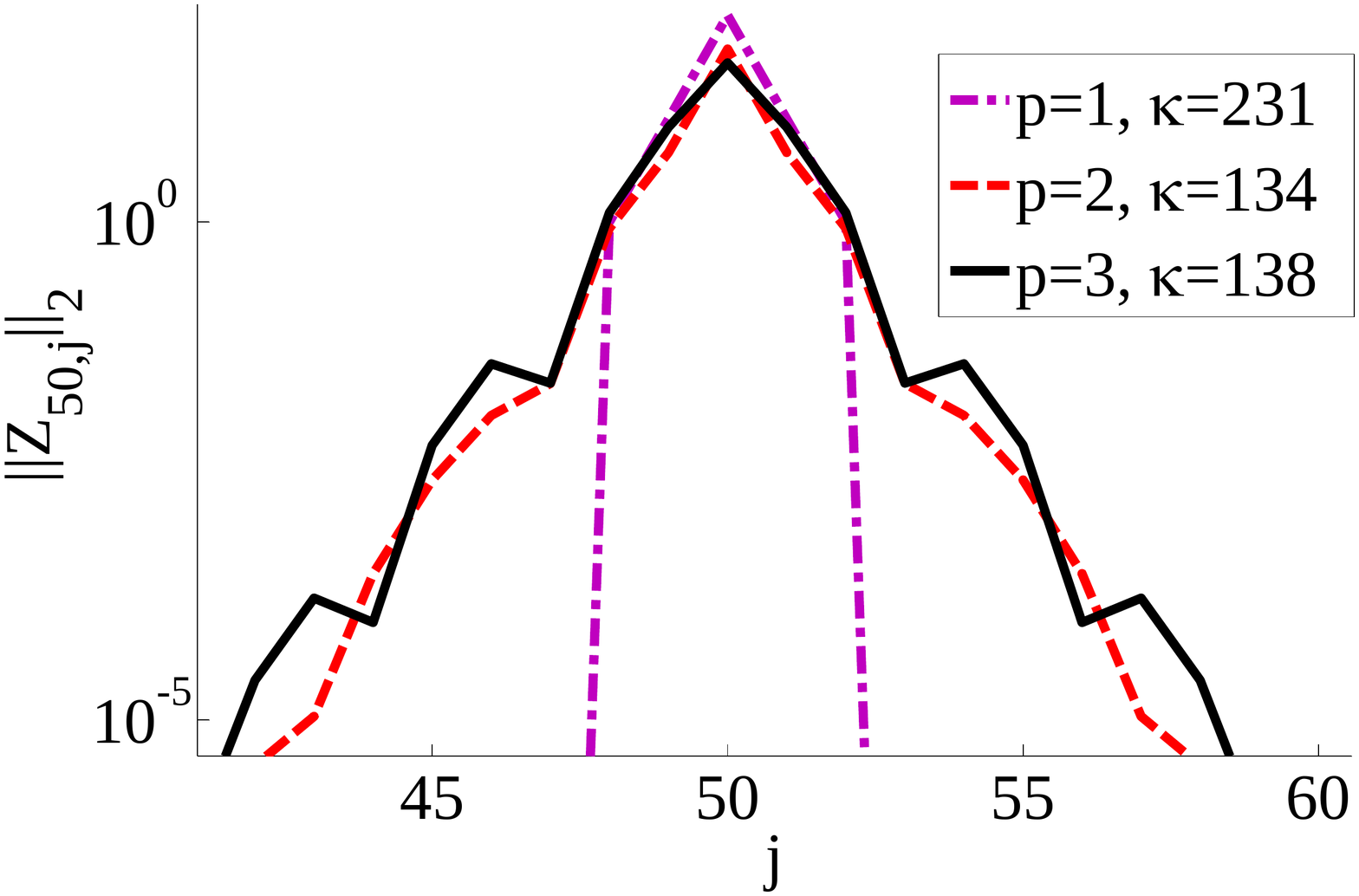}}
 \subfloat[]{\includegraphics[width=4.2cm, height=3.9cm, trim=0mm 0mm 0mm 0mm ,clip=true]{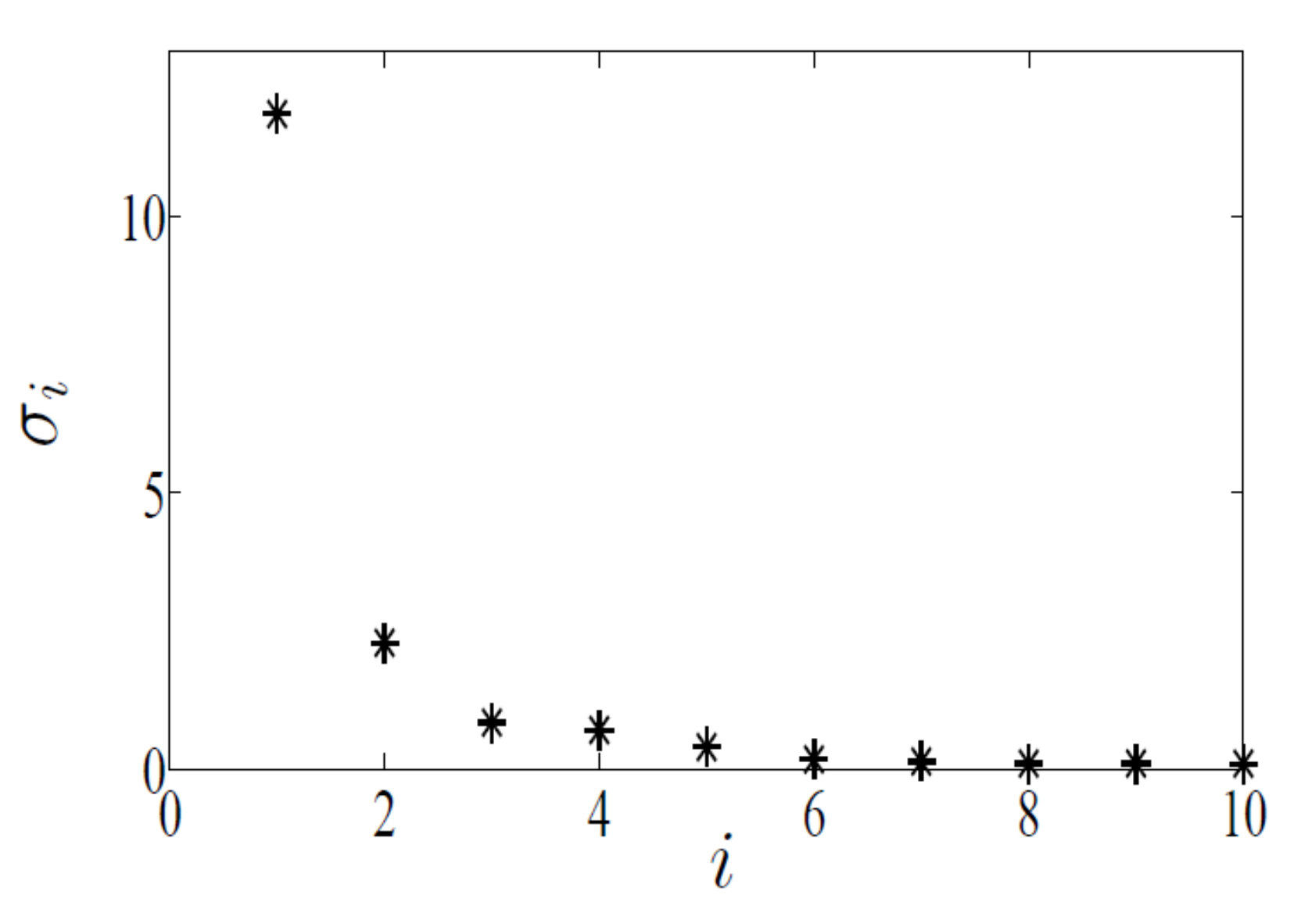}}\;\;
 \caption{\small{(a) The norm of the block elements $Z_{50,j}\in \mathbb{R}^{2\times 2}$ of the $50$th block row of $\mathcal{D}$. (b) The singular values of the data matrix used to determine the order and to estimate the state-sequence of $\mathcal{S}_{2}$. The data matrix is formed on the basis of the input $3$ in \eqref{inputSelections}.}}
\label{decayFinv}
\end{figure}

Figure \ref{decayFinv}(a) shows how the off-diagonal decay of $\mathcal{D}$ depends on $p$ and $\kappa$. The results presented in Fig. \ref{decayFinv}(a) confirm that for well-conditioned $\mathcal{J}_{2p}$, the off-diagonal decay of $\mathcal{D}$ is rapid. This figure also suggests that the accuracy of approximating $\mathcal{D}$ by $\breve{\mathcal{D}}_{t}$, is relatively good for $t=1$.
 To illustrate how the quality of the identified model depends on the selection of the input vector of \eqref{unifiedSys}, we identify the state-sequence of $\mathcal{S}_{2}$ using 5 different inputs:
\begin{small}
\begin{align}
&1. \;\breve{\Omega}_{2}=\mathbf{u}_{2}(k) \notag 
 \\
&2. \;\breve{\Omega}_{2}=\text{col}\left(\mathbf{y}_{1}(k),\mathbf{y}_{2}(k),\mathbf{y}_{3}(k),\mathbf{u}_{2}(k)\right) \notag 
 \\
&3. \;\breve{\Omega}_{2}=\text{col}\left(\mathcal{Y}_{1,k-1}^{k},...,\mathcal{Y}_{3,k-1}^{k},\mathbf{u}_{2}(k),\mathbf{u}_{1}(k-1),..., \mathbf{u}_{3}(k-1)\right) \notag 
 \\
&4. \;\breve{\Omega}_{2}= \notag \\
&\text{col}\left(\mathbf{y}_{1}(k-1),...,\mathbf{y}_{6}(k-1),\mathbf{u}_{2}(k),\mathbf{u}_{1}(k-1),..., \mathbf{u}_{6}(k-1)\right) 
\notag \\
& 5. \;\breve{\Omega}_{2}=\text{col}(\mathcal{Y}_{1,k-1}^{k},...,\mathcal{Y}_{6,k-1}^{k},\mathbf{u}_{2}(k),\mathbf{u}_{1}(k-1),..., \mathbf{u}_{6}(k-1)) 
\label{inputSelections}
\end{align}
\end{small}
In the case of inputs $3,4$ and $5$, data matrices used to estimate the Markov parameters (impulse response parameters) of \eqref{unifiedSys} are ill conditioned. This ill-conditioning is caused by the fact that the local outputs, that are the elements of $\breve{\Omega}_{2}$, are depending on delayed outputs and inputs. We use the regularization technique to improve the condition number of a data matrix used for identification of the Markov parameters of  $\mathcal{S}_{2}$ (the regularization parameter is $0.05$). Local state order is selected by examining the singular values of the data matrix that is formed on the basis of $\{\breve{\Omega}_{2}, \mathbf{y}_{2}(k)\}$. For each of the inputs defined in \eqref{inputSelections}, we form the data matrix and we select the local order $n=2$. For illustration, in Fig. \ref{decayFinv}(b) we present the singular values of the data matrix formed on the basis of the input 3 (similar behavior of singular values can be observed for the inputs $2$, $4$ and $5$, while in the case of the input $1$ the state order could not be uniquely determined). 
\\
Using the similar procedure, we determine the order and we estimate the state sequence of $\mathcal{S}_{1}$. The state sequence of $\mathcal{S}_{1}$, can be also identified on the basis of \eqref{localSubSys} (where $i$ is set to 1). Namely, using $\{\hat{\mathbf{x}}_{2}(k),\mathbf{u}_{1}(k) \}$ as known inputs in \eqref{localSubSys}, and using $\{\mathbf{y}_{1}(k) \}$, we can directly identify $\{\hat{\mathbf{x}}_{1}(k)\}$ by solving a simple least-squares problem. \\
Next, we estimate the local system matrices $\{\hat{A},\hat{E},\hat{B},\hat{C}\}$. Using $\{\hat{A},\hat{E},\hat{B},\hat{C}\}$ we form the estimates of the global system matrices $\{\underline{\hat{A}},\underline{\hat{B}},\underline{\hat{C}}\}$ and we compute the VAF of the global model. The average values of VAF (for $\mathcal{S}_{2}$) are presented in Table \ref{tab:VAF}.
\par
\begin{figure}[H]
\centering
 \subfloat[]{\includegraphics[width=4.1cm, height=3.5cm,  trim=0mm 0mm 0mm 0mm ,clip=true]{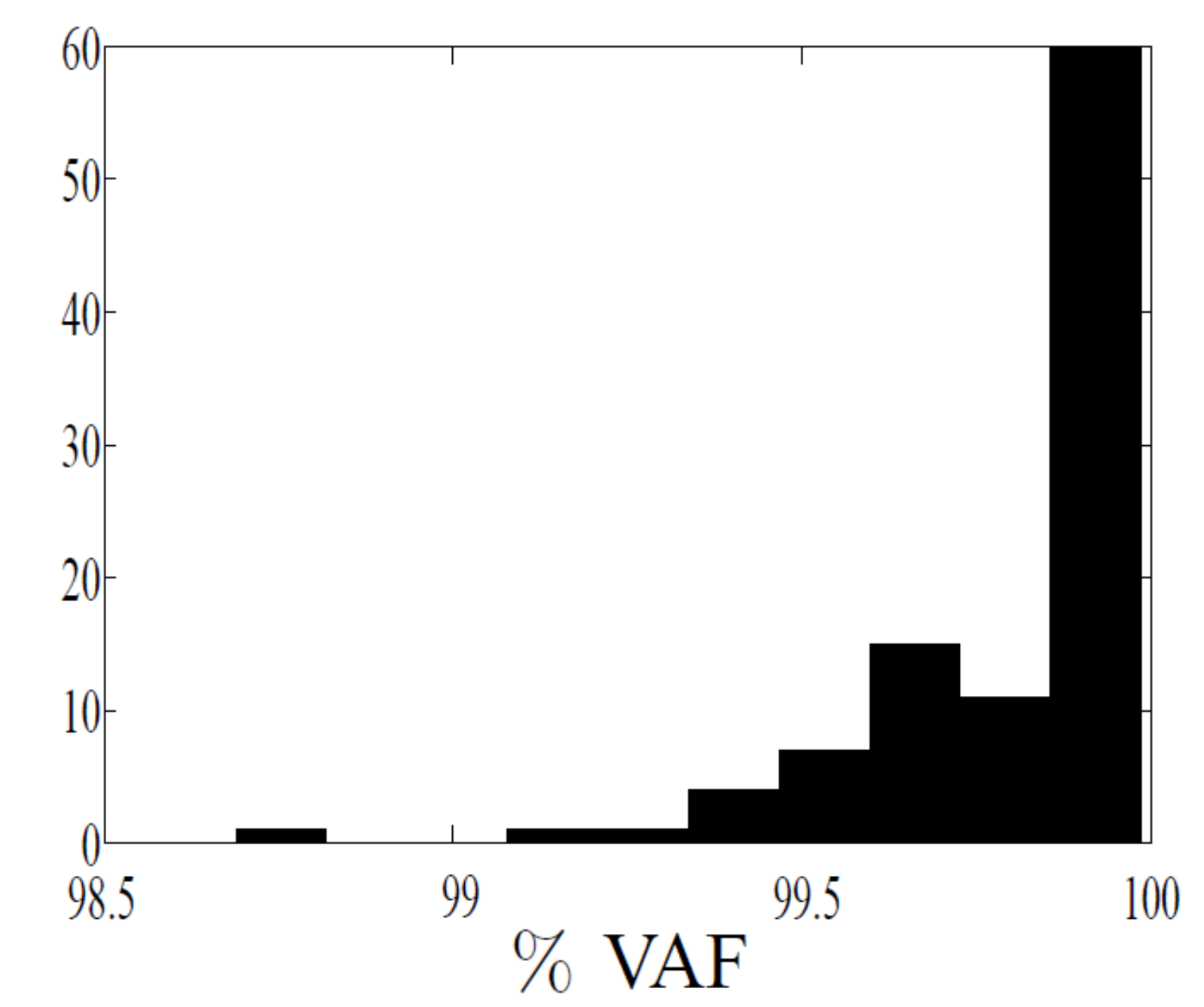}}\;\;
 \subfloat[]{\includegraphics[width=4.1cm, height=3.5cm, trim=0mm 0mm 0mm 0mm,clip=true]{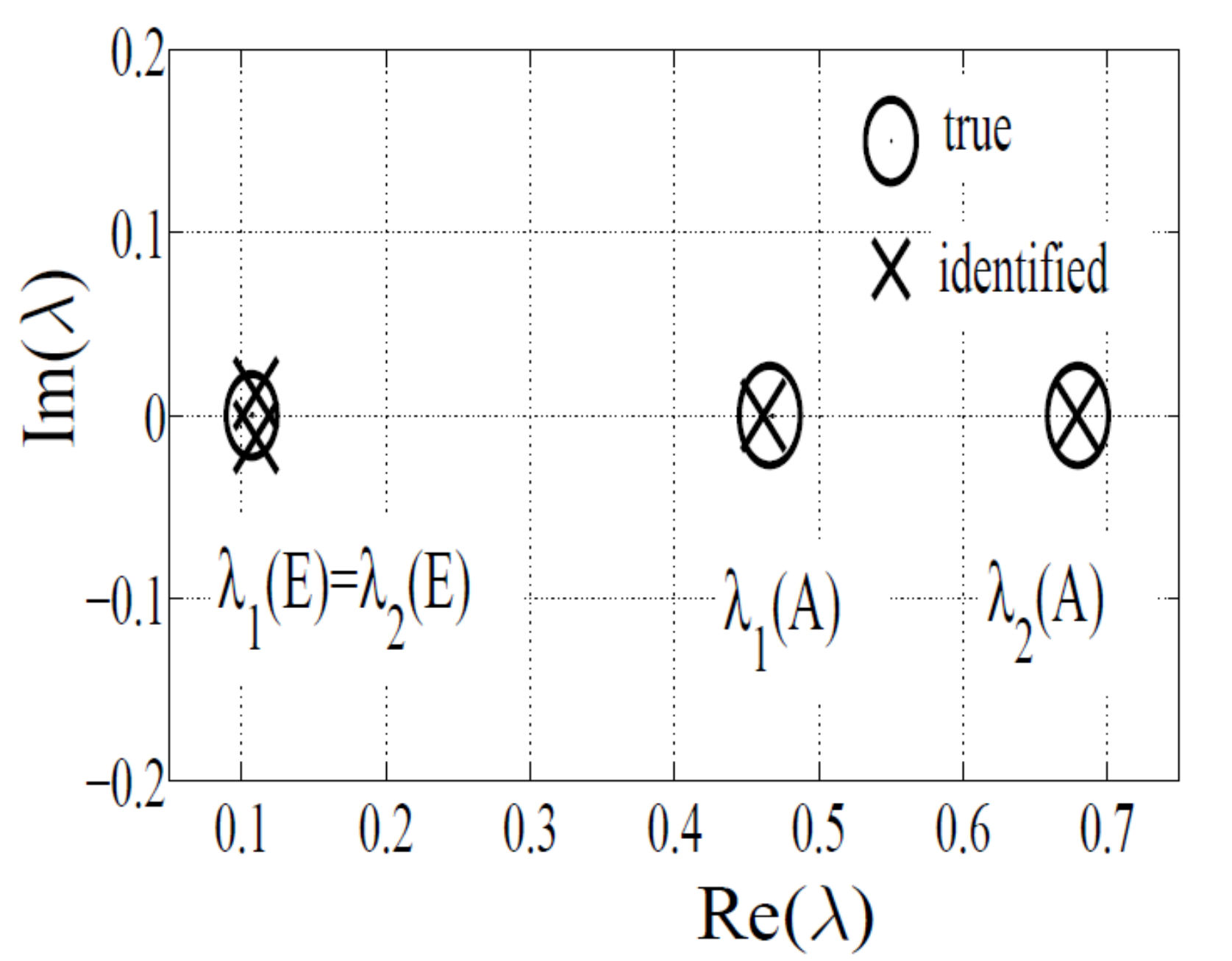}}\;\;
 \caption{\small{(a) Distribution of the VAF of $\mathcal{S}_{1}$. The identification of the state-sequence of $\mathcal{S}_{2}$ is performed using input 2, defined in \eqref{inputSelections}. (b) Eigenvalues of the estimated matrix $\hat{A}$ and $\hat{E}$, when the input 2 is used for identification. The outputs are not corrupted by noise.}}
\label{fig:Nonoise}
\end{figure}
\begin{figure}[H]
\centering
 \subfloat[]{\includegraphics[width=4.1cm, height=3.5cm,  trim=0mm 0mm 0mm 0mm ,clip=true]{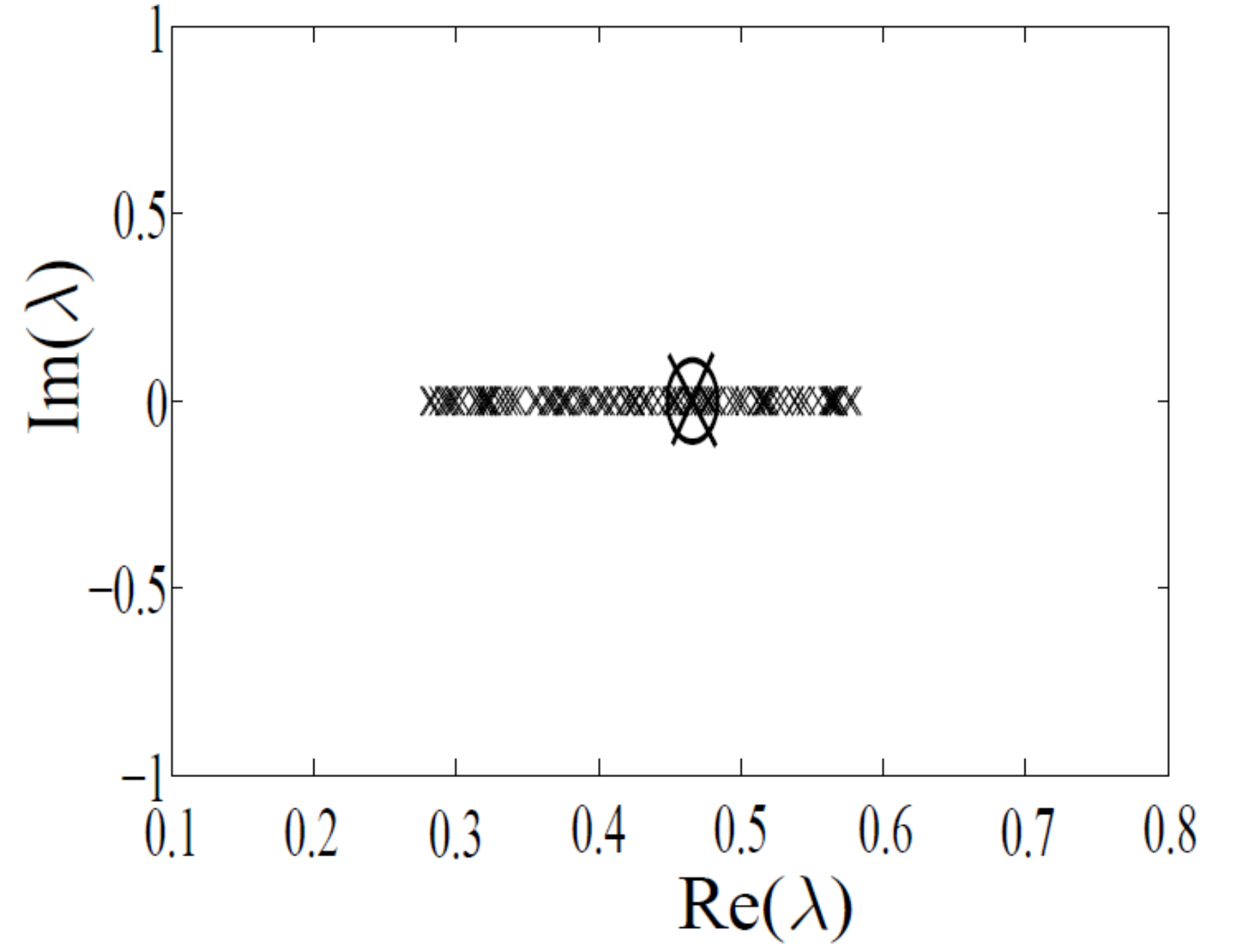}}\;\;
 \subfloat[]{\includegraphics[width=4.1cm, height=3.5cm, trim=0mm 0mm 0mm 0mm,clip=true]{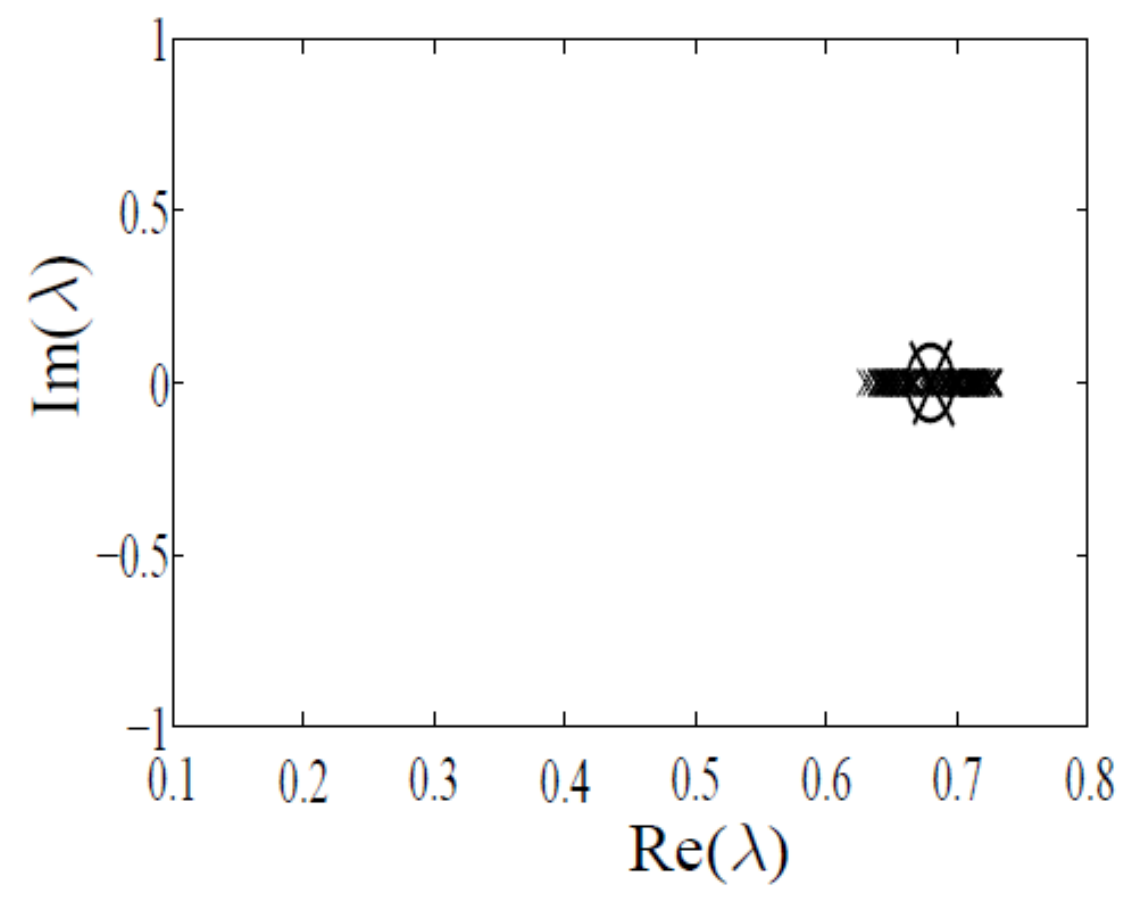}}\;\;
 \caption{\small{(a) and (b): The distribution of the eigenvalues of $\hat{A}$ for 100 identification experiments. The SNR ratio is $25$ $[\text{dB}]$. The circle with the big "X" corresponds to the eigenvalue of $A$. Identification of the state-sequence of $\mathcal{S}_{2}$ is performed using input 2, defined in \eqref{inputSelections}.}}
\label{eigenvalues}
\end{figure}

\begin{tiny}\begin{table}[H]
\centering
\begin{tabular}{ |c | c | c | c |c|c|}
\hline
input                              &    1               &        2                &             3            &               4          &     5\\
\hline
VAF  (without reg.)                              &   40 \%          &         99.7 \%    &       5 \%             &           30 \%      &      20 \%  \\
\hline      
VAF (with reg.)        &     -              &          99.6 \%       &        97.7 \%         &         99.2 \%     &      98.5 \% \\
\hline
\end{tabular}
\caption{\small{The average values of VAF for 5 different inputs (reg. is the abbreviation for regularization).}}
\label{tab:VAF}
\end{table}\end{tiny}
From Table \ref{tab:VAF} we conclude that "best" identification results are obtained when the input 2 is used for identification. This input is formed by eliminating the delayed inputs and outputs that cause ill-conditioning. In Fig. \ref{fig:Nonoise}(a), we present the distribution of VAF values for the output of $\mathcal{S}_{1}$, when the input 2 is used for identification (in total $100$ identification experiments are performed). Similar results are obtained for other local subsystems. The eigenvalues of $\hat{A}$, when the input 2 is used for identification, are given in Fig. \ref{eigenvalues}. Next, assuming that the local outputs are not corrupted by the noise, we perform identification using the input 2 (with regularization). The results are given in Fig. \ref{fig:Nonoise}(b).
As it can be seen from Fig. \ref{fig:Nonoise}(b), in the noise-free scenario we are able to obtain a relatively good identification results. Some of the eigenvalues are biased. This is mainly because of the approximation errors in the state-space model \eqref{unifiedSys} and because a part of useful information is "thrown away" by forming the input 2 (that is, some information is lost by eliminating the delayed inputs and outputs).
\section{Conclusion}
\label{sectionConclusion}
In this paper we proposed a decentralized subspace algorithm for identification of state-space models of large-scale interconnected systems. To develop the identification algorithm, we proved that the state of a local subsystem can be approximated by a linear combination of inputs and outputs of local subsystems that are in its neighborhood. For systems with well-conditioned, finite-time observability Gramians, the size of this neighborhood is small. Consequently, the local subsystems can be estimated in a computationally efficient manner. The numerical experiments confirm the effectiveness of the proposed algorithm. 
\bibliographystyle{unsrt}
\bibliography{bibl}

\end{document}